\documentclass[twoside,leqno,twocolumn]{article}
\usepackage{ltexpprt}
\usepackage{graphicx}
\usepackage{bm,bbm}
\usepackage{hyperref}
\usepackage[margin=2cm]{geometry}
\usepackage{xcolor}

\usepackage{amsmath}
\usepackage{amssymb}
\usepackage{tikz}
\usepackage{subcaption}

\newcommand{\per}{\text{per}}
\newcommand{\haf}{\text{haf}}
\newcommand{\lhaf}{\text{lhaf}}
\newcommand{\bs}{Boson Sampling }
\newcommand{\gbs}{Gaussian Boson Sampling }

\begin{document}

\title{A faster hafnian formula for complex matrices \\ and its benchmarking on a supercomputer}
\author{Andreas Bj\"{o}rklund \thanks{Dept. Comp. Sci., Lund University, Sweden} \\
\and
Brajesh Gupt \thanks{Xanadu, Toronto, Canada}
\and 
Nicol\'as Quesada $^{\dagger}$
}
\date{}

\maketitle






\pagenumbering{arabic}
\setcounter{page}{1}

\begin{abstract} \small\baselineskip=9pt 
We introduce new and simple algorithms for the calculation of the number of perfect matchings of complex weighted, undirected graphs with and without loops.  Our compact formulas for the hafnian and loop hafnian of $n \times n $ complex matrices run in $O(n^3 2^{n/2})$ time, are embarrassingly parallelizable and, to the best of our knowledge, are the fastest exact algorithms to compute these quantities. 
Despite our highly optimized algorithm, numerical benchmarks on the Titan supercomputer with matrices up to size $56 \times 56$ indicate that one would require the 288000 CPUs  of this machine for about a month and a half to compute the hafnian of a $100 \times 100$ matrix.
\end{abstract}

\section{Introduction}
Counting perfect matchings in a graph is an important problem in graph theory\cite{barvinok2016combinatorics} and has diverse applications \cite{hosoya1971topological,quesada2018faster}. For a bipartite graph the number of perfect matchings is given by the permanent of the associated adjacency matrix, which has been shown to be \#P-complete to compute exactly \cite{valiant1979complexity}. Various algorithms have been developed for the fast computation of permanents  \cite{ryser1963combinatorial,BaxFranklin,Balasubramanian} (see Ref. \cite{wu2016computing} for a recent and detailed benchmarking of different algorithms for the computation of permanents using supercomputers). For a more general graph (one that is not bipartite), the number of perfect matchings is given by the \emph{hafnian} of the associated adjacency matrix of the graph \cite{BarvinokIntro}. The hafnian can be thought of as a generalization of the permanent. Whereas the permanent counts the number of perfect matchings in a bipartite graph, the hafnian counts the number of perfect matchings in an undirected graph.
For the related problem of \emph{approximating} the hafnian several methods have been developed for restricted sets of matrices  \cite{2016arXiv160107518B, 2014arXiv1409.3905R, Chien:2004:DAC:982792.982903,10.1007/3-540-36494-3_38}.

In this manuscript we develop a new algorithm to compute hafnians of general complex matrices that runs in $O(n^3 2^{n/2})$ time where $n$ is the size of the matrix. Our algorithm builds on the hafnian algorithm of Cygan and Pilipczuk \cite{cygan2015faster}, here adapted to the field of complex numbers using elementary tools from linear algebra. Compared to the general ring hafnian algorithm of Cygan and Pilipczuk, our algorithm is a factor $n$ faster. {This factor of $n$ is gained by replacing a dynamic programming tabulation running in $O(n^4)$ operations in a ring by a Schur decomposition of real or complex matrices which can be done in $O(n^3)$ operations.}
The new algorithm is, to the best of our knowledge, the fastest exact algorithm to compute the hafnian of a complex matrix. 

A second motivation for studying fast computation of hafnians stems from quantum computing. Recent developments in quantum complexity theory have provided renewed impetus for the study of combinatorial sampling problems. The most prominent of these developments is perhaps Aaronson and Arkhipov's \bs problem \cite{aaronson2011computational}. In \bs $n$ photons are sent through a (linear) lossless optical device that has $n^2$ inputs and outputs. 
The probability that a certain arrangement of detectors click is proportional to $|\text{per}(U_{S})|^2$, where $U_{S}$ is a submatrix of a unitary matrix $U$ representing the optical device and `per' stands for permanent \cite{ryser1963combinatorial}. Aaronson and Arkhipov argue that for large enough $n$ it will be impossible for a classical computer to generate samples in polynomial time from the event distribution (of click detections) of the optical circuit just described. 
{Of course, the number of photons $n$ and interferometer size $n^2$ after which  classical methods and hardware cannot keep up with quantum hardware will depend not only on the quantum hardware, but also on how far  classical algorithms can be pushed\cite{wu2016computing,neville2017classical,clifford2018classical}.}
{It is well understood that these classical algorithms should scale exponentially in $n$\cite{aaronson2011computational,clifford2018classical}, yet to delineate the boundary of quantum supremacy it is necessary to limit as much as possible the polynomial prefactors that accompany these exponentials. For instance in the work of Neville \emph{et al.}\cite{neville2017classical} Ryser's formula\cite{ryser1963combinatorial} with Gray code ordering is used to calculate permanents that are fed into a Metropolis independent sampling Markov Chain Montecarlo to generate boson samples.
}


Recently Hamilton et al. \cite{hamilton2017gaussian,kruse2018detailed} introduced a related problem called Gaussian Boson Sampling. Their problem is almost identical to \bs except that now the light sent into the optical device is not single photons but squeezed light \cite{lvovsky2014squeezed}. Hamilton et al. show that this ``small'' change can significantly simplify the experimental challenges in constructing a Boson Sampler.
In \gbs the probability of the detectors clicking is now proportional to the modulus squared of the hafnian \cite{barvinok2016combinatorics} of a \emph{complex} full-rank submatrix constructed from the unitary matrix representing the circuit and the values of the intensities of the squeezed light going into the device {(In appendix \ref{app:lowrank} we study in detail the dependence on the rank for low rank matrices using methods developed by Barvinok \cite{B96}). Finally, note that the approximate methods developed for counting perfect matchings are aimed at (weighted-)graphs with real or positive entries  \cite{2016arXiv160107518B, 2014arXiv1409.3905R,10.1007/3-540-36494-3_38} making them unsuitable for GBS.}

With the development of new quantum sampling problems that are less complex to implement experimentally it becomes important to understand where the limits of classical computers with the best possible algorithms lie \cite{harrow2017quantum}. Thus, the results presented here should be of relevance for any claim of quantum supremacy using Gaussian Boson Sampling. Moreover, for the case of boson sampling, the polynomial prefactors appearing in the complexity of the hafnian calculation will play an important role in determining where quantum supremacy lies for Gaussian Boson Sampling\cite{quesada2018gaussian,gupt2018classical}.

\subsection{Earlier Work}
The exact calculation of the number of perfect matchings for general graphs has been investigated by several authors in recent years. An algorithm running in $O(n^2 2^n)$  time was given by Bj\"{o}rklund and Husfeldt \cite{bjorklund2008exact} in 2008. {In the same paper an algorithm running in $O(1.733^n)$ time was presented using fast matrix multiplication. In the same year Kan\cite{kan2008moments} presented an algorithm, for positive definite real matrices, running in time $O(n 2^n)$ by representing the hafnian as a moment of the multinormal distribution.} Koivisto \cite{koivisto2009partitioning} gave an $O^*(\phi^n)$ time and space algorithm, where $\phi = (1+\sqrt{5})/2 \approx 1.618$ is the Golden ratio and the notation $O^*$ is used to indicate that polylogarithmic corrections have been suppressed in the scaling. Nederlof \cite{nederlof2009fast} provided a polynomial space algorithm running in $O(1.942^n)$ time.

Finally, Bj\"{o}rklund \cite{bjorklund2012counting} and later Cygan and Pilipczuk \cite{cygan2015faster} provided $O(\text{poly}(n) 2^{n/2})$ time and polynomial space algorithms for the calculation of the general ring hafnian. These algorithms are believed to be close to optimal unless there are surprisingly efficient algorithms for the Permanent. This is because these two algorithms can also be used to count (up to polynomial corrections) the number of perfect matchings for bipartite graphs with the same exponential growth as Ryser's algorithm for the permanent \cite{ryser1963combinatorial}. Equivalently, if one could construct an algorithm that calculates hafnians in time $O(\alpha^{n/2})$ with $\alpha<2$ one could calculate permanents \emph{faster} than Ryser's algorithm (which is the fastest known algorithm to calculate the permanent \cite{rempala2007symmetric}). This is because of the identity
\begin{align}\label{eq:bipartite}
\haf \left( \left[
\begin{array}{cc}
0 & \bm{W} \\
\bm{W}^T & 0 \\
\end{array}
\right]\right) = \per(\bm{W}), 
\end{align}
which states that a bipartite graph with two parts having $n/2$ elements can always be thought as a simple graph with $n$ vertices.
It should be noted that improving over Ryser's algorithm is a well-known open problem: e.g. Knuth~\cite{TAOCP} asks for an arithmetic circuit for the permanent with less than $2^n$ operations. Also note that since the exact calculation of the permanent of (0,1) matrices is in the \#P complete class \cite{valiant1979complexity} the above identity shows that deciding if the hafnian of a complex matrix is larger than a given value is also in the \#P complete class.

\subsection{Our Contribution}
In this paper we improve upon recently developed algorithms for counting the number of perfect matchings of undirected graphs \cite{bjorklund2012counting,cygan2015faster} and the calculation of hafnians.
Furthermore these algorithms are generalized to allow for the inclusion of graphs that contain loops {which have been recently shown to be linked to the calculation of spectral lines of molecules\cite{quesada2018faster}}. Finally, we provide benchmarks of the algorithms developed using the Titan supercomputer of the Oak Ridge National Laboratory. The results presented here should provide a stepping stone for our understanding of how fast can Hafnians be calculated in classical computers and delimit the realm of quantum supremacy for (Gaussian) Boson Samplers.

\section{Hafnians and Perfect Matchings}
The hafnian of an $n \times n $ symmetric matrix $\bm{A} =\bm{A}^T$ is defined as
\begin{align}
\haf(\bm{A}) = \sum_{M \in \text{PMP}(n)} \prod_{\scriptscriptstyle (i, j) \in M} A_{i, j},
\label{eq:hafA}
\end{align}
where PMP$(n)$ stands for the set of perfect matching permutations of $n$ (even) objects (see \ref{app:lowrank} for more details on the set PMP$(n)$). 
For $n=4$ the set of perfect matchings is 
\begin{align}\label{eq:PMP4}
\text{PMP}(4) = \big\{ (0,1)(2,3),\ (0,2)(1,3),\ (0,3)(1,2) \big\},
\end{align}
and the hafnian of a $4 \times 4$ matrix $\bm{B}$ is
\begin{align}\label{eq:hafB}
\haf(\bm{B}) = B_{0,1} B_{2,3}+B_{0,2}B_{1,3}+B_{0,3} B_{1,2}.
\end{align}
More generally, the set PMP($n$) contains 
\begin{align}\label{eq:haf1}
|\text{PMP}(n)|=(n -1)!! = 1 \times 3 \times 5 \times \ldots \times (n -1),
\end{align}
elements and thus as defined it takes $(n-1)!!$ additions of products of $n/2$ numbers to calculate the hafnian of $\bm{A}$. 
Note that the diagonal elements of the matrix $\bm{A}$ do not appear in the calculation of the hafnian and are (conventionally) taken to be zero. 

\begin{figure}[!ht]
  \input{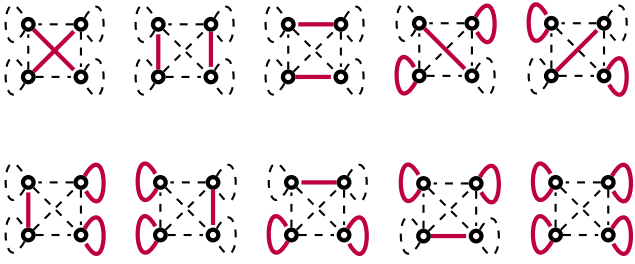}
  \caption{\label{fig:single-pair} Single-pair matchings for a complete 4 vertex graph. 
    The first three partitions in the top row correspond to the set PMP$(4)$.}
\end{figure}

The hafnian function has an interesting connection with graph theory: if $\bm{A}$ is the adjacency matrix of a loopless, unweighted, undirected graph (i.e., $\bm{A}$ is a (0,1) matrix with zeros along the diagonal) then haf($\bm{A}$) is precisely the number of perfect matchings of the graph represented by $\bm{A}$. 
{A matching is a subset of the edges of a graph in which no two edges share a vertex. A perfect matching is a matching which matches all the vertices of the graph. }

Each element of the set of perfect matchings asserts whether the partition of the graph leaves no edge unmatched. For the set PMP(4) in Eq. (\ref{eq:PMP4}) one can easily visualize the corresponding matchings of a graph with four vertices as the first three partitions in the top row of Fig. \ref{fig:single-pair}.
The notion of perfect matching is easily generalized from (0,1) adjacency matrices to matrices over any field. 
The hafnian of a (symmetric) matrix $\bm{A}$ will be then the sum of weighted perfect matchings of an undirected graph with adjacency matrix $\bm{A}$.

In this manuscript we will also study a generalization of the hafnian function where we will consider graphs that have loops, henceforth referred to as `lhaf' (loop hafnian). The weight associated with said loops will be allocated in the diagonal elements of the adjacency matrix $\bm{A}$ (which were previously ignored in the definition of the hafnian in Eqs. (\ref{eq:hafA}) and (\ref{eq:hafB})). To account for the possibility of loops we generalize the set of perfect matching permutations PMP to the single-pair matchings (SPM). This is simply the set of perfect matchings of a complete graph with loops. Thus we define
\begin{align}
\lhaf(\bm{A}) = \sum_{M \in \text{SPM}(n)} \prod_{\scriptscriptstyle (i,j) \in M} A_{i,j}.
\end{align}
Considering again a graph with 4 vertices we get a total of 10 SPMs:
\begin{align}\label{eq:SPM4}
&\text{SPM}(4)=\big\{ (0,1)(2,3),\ (0,2)(1,3), \ (0,3)(1,2),  \\
& \ (0,0)(1,1)(2,3),\ (0,1)(2,2)(3,3), \ (0,2)(1,1)(3,3),  \nonumber\\
& \ (0,0)(2,2)(1,3), \ (0,0)(3,3)(1,2),\ (0,3)(1,1)(2,2), \nonumber\\ 
& \ (0,0)(1,1)(2,2)(3,3) \big\}. \nonumber
\end{align}
and the lhaf of a $4 \times 4$ matrix $\bm{B}$ is
\begin{align}
\lhaf(\bm{B}) =& B_{0,1} B_{2,3}+B_{0,2}B_{1,3}+B_{0,3} B_{1,2}\\
&+ B_{0,0} B_{1,1} B_{2,3}+B_{0,1} B_{2,2} B_{3,3}+B_{0,2}B_{1,1}B_{3,3}\nonumber\\
&+ B_{0,0} B_{2,2} B_{1,3}+B_{0,0}B_{3,3}B_{1,2}+B_{0,3} B_{1,1} B_{2,2}\nonumber\\
&+ B_{0,0} B_{1,1} B_{2,2} B_{3,3}. \nonumber
\end{align}
More generally for a graph with $n$ vertices ($n$ even) the number of SPMs is 
\begin{align}\label{eq:haf2}
|\text{SPM}(n)| = T\left(n,-\tfrac{1}{2},\tfrac{1}{\sqrt{2}}\right) = T(n)
\end{align}
where $T(a,b,r)$ is the Toronto function (cf. page 509 of Ref. \cite{abramowitz1964handbook}) and 
where $T(n)$ is the $n^{\text{th}}$ telephone number.
A derivation of this formula and some comments on the asymptotic super polynomial scaling of the ratio between the number of perfect matching and the number of single-pair matchings are presented in Appendix \ref{countingSPM}. 
Note that asymptotically 
\begin{align}
\frac{|\text{SPM}(n)|}{|\text{PMP}(n)|}  = \frac{T\left(n \right)}{(n-1)!!} \sim \exp(\sqrt{n})
\end{align}

Finally, let us comment on the scaling properties of the $\haf$ and $\lhaf$.
Unlike the hafnian the loop hafnian function is not homogeneous in its matrix entries, i.e.
\begin{align}
\haf(\mu \bm{A}) &= \mu ^{n/2} \haf(\bm{A}) \text{  but},\\
\lhaf(\mu \bm{A}) &\neq \mu ^{n/2} \lhaf(\bm{A}).
\end{align}
where $n$ is the size of the matrix $\bm{A}$ and $\mu \geq 0$. However if we split the matrix $\bm{A}$  in terms of its diagonal $\bm{A}_{\text{diag}}$ part and its offdiagonal part $\bm{A}_{\text{off-diag}}$
\begin{align}
\bm{A} = \bm{A}_{\text{diag}}+\bm{A}_{\text{off-diag}},
\end{align}
then it holds that 
\begin{align}
\label{scaling}
\lhaf(\sqrt{\mu} \bm{A}_{\text{diag}}+ \mu \bm{A}_{\text{off-diag}}) &= \mu^{n/2} \lhaf(\bm{A}_{\text{diag}}+ \bm{A}_{\text{off-diag}}) \nonumber \\
&=\mu^{n/2} \lhaf(\bm{A}).
\end{align}
Later we will show that the new formulas we derive here for the hafnian and loop hafnian explicitly respect these scaling relations.
Finally, note that when all the diagonal elements of the matrix $\bm{A}$ are set to 1, the loop hafnian function can be used to count the number of matchings of a loopless graph with adjacency matrix $\bm{A}$.

\section{The Algorithm}
As mentioned in the previous section, the hafnian and loop hafnian functions count the number of perfect matchings in a graph. The topology of the graph is encoded in the adjacency matrix that is input into either function. In the following sections we present an algorithm that allows us to count the number of perfect matchings of a graph with $n$ vertices in time $O(n^3 2^{n/2})$ for (unweighted) graphs with and without loops, and then generalize it to weighted graphs. Our algorithm and its analysis largely follows that of Cygan and Pilipczuk~\cite{cygan2015faster} with one crucial exception. Whereas the terms in their formula are computed by an $O(n^4)$ dynamic programming tabulation, we reduce the terms to efficiently computable functions of the traces of the first $n/2$ powers of a matrix. This enables us to gain a factor of $n$ in the running time by first computing the eigenvalue spectrum with a known $O(n^3)$ time algorithm. We can then use standard trace identities to compute all traces of the matrix powers more efficiently than by explicitly constructing the matrix powers.

\subsection{Notation and Terminology}
 
Let $G=(V,E)$ be an undirected graph with loops, and let $n=|V|$ be even. A \emph{perfect matching} in $G$ is a subset $E'\subset E$ of edges such that every vertex in $V$ is part of exactly one edge $e\in E'$. Note again that $e$ may be a loop from a vertex $v$ to itself. We consider here the problem of enumerating all possible perfect matchings of $G$, a quantity we will denote by $M(G)$.

We write $[m]$ for a positive integer $m$ as the set 
\begin{align}
[m] =\{0,1,\ldots,m-1\}. 
\end{align}
The vertices $V$ of the graph will be associated with the set $[n]$.
A \emph{walk} is a sequence of vertices $\hat{w}=(w_0,w_1,\ldots,w_\ell)$ where $\forall i<\ell:(w_i,w_{i+1})\in E$. The length of the walk is $\ell$. 

For a subset $A\subseteq E$, we say a walk $\hat{w}=(w_0,w_1,\ldots,w_\ell)$ is \emph{$A$-alternating}, if and only if 
\begin{itemize}
\item $\forall i<\ell, i \mbox{ is odd }: (w_i,w_{i+1})\in A$,
\item $\forall i<\ell, i \mbox{ is even }: (w_i,w_{i+1})\not \in A$, and
\item either $\ell$ is even and $\hat{w}$ is closed ($w_0=w_\ell$), or $\ell$ is odd and the endpoints are loops ($w_0=w_1$ and $w_{\ell-1}=w_\ell$).
\end{itemize}
An \emph{$A$-tangle} is a set of $A$-alternating walks passing through edges in $A$ exactly $n/2$ times in total, possibly by traversing some edges several times.

\subsection{Perfect matchings in exponential time}
\begin{figure}[!t]
\begin{subfigure}[b]{0.12\textwidth}
\begin{tikzpicture}[scale=.3, shorten >=1pt, auto, node distance=1cm, ultra thick]
    \tikzstyle{node_style} = [circle,draw=black, inner sep=0pt, minimum size=4pt]
    \tikzstyle{edge_style} = [-,draw=black, line width=2, ultra thick]
    \tikzstyle{loop_style_right} = [-,draw=black, line width=2, ultra thick,in=-60,out=60,looseness=10]

    \node[node_style] (v1) at (0,0) {};
    \node[node_style] (v2) at (-2,2) {};
    \node[node_style] (v3) at (0,2) {};
    \node[node_style] (v4) at (2,2) {};
    \node[node_style] (v5) at (-3,4) {};
    \node[node_style] (v6) at (-1,4) {};
    \node[node_style] (v7) at (1,4) {};
    \node[node_style] (v8) at (3,4) {};
    \node[node_style] (v9) at (-1,6) {};
    \node[node_style] (v10) at (1,6) {};

    \draw[edge_style]  (v1) edge (v2);
    \draw[edge_style]  (v1) edge (v3);
    \draw[edge_style]  (v1) edge (v4);
    \draw[edge_style]  (v2) edge (v3);
    \draw[edge_style]  (v3) edge (v4);
    \draw[edge_style]  (v2) edge (v5);
    \draw[edge_style]  (v4) edge (v8);
    \draw[edge_style]  (v5) edge (v6);
    \draw[edge_style]  (v6) edge (v7);
    \draw[edge_style]  (v7) edge (v8);
    \draw[edge_style]  (v5) edge (v9);

    \draw[edge_style]  (v6) edge (v9);
    \draw[edge_style]  (v7) edge (v10);
    \draw[edge_style]  (v8) edge (v10);
    \draw[edge_style]  (v9) edge (v10);
    \draw[edge_style]  (v5) to [loop left, in=120,out=240, looseness=10] (v5);
    \draw[edge_style]  (v8) to [loop right, in=-60,out=60,looseness=10 ] (v8);
    \end{tikzpicture}
\end{subfigure}
\quad \quad 
\begin{subfigure}[b]{0.12\textwidth}
    \begin{tikzpicture}[scale=.3, shorten >=1pt, auto, node distance=1cm, ultra thick]
    \tikzstyle{node_style} = [circle,draw=black, inner sep=0pt, minimum size=4pt]
    \tikzstyle{edge_style} = [draw=black, line width=2, ultra thick]
    \tikzstyle{edge_styler} = [-,draw=red, line width=2, ultra thick, dashed]
    \tikzstyle{edge_styleg} = [-,draw=red, line width=2, ultra thick, dotted]
    
    \node[node_style] (v1) at (0,0) {};
    \node[node_style] (v2) at (-2,2) {};
    \node[node_style] (v3) at (0,2) {};
    \node[node_style] (v4) at (2,2) {};
    \node[node_style] (v5) at (-3,4) {};
    \node[node_style] (v6) at (-1,4) {};
    \node[node_style] (v7) at (1,4) {};
    \node[node_style] (v8) at (3,4) {};
    \node[node_style] (v9) at (-1,6) {};
    \node[node_style] (v10) at (1,6) {};

    \draw[edge_style]  (v1) edge (v2);
    \draw[edge_style]  (v1) edge (v3);
    \draw[edge_style]  (v1) edge (v4);
    \draw[edge_style]  (v2) edge (v3);
    \draw[edge_style]  (v3) edge (v4);
    \draw[edge_style]  (v2) edge (v5);
    \draw[edge_style]  (v4) edge (v8);
    \draw[edge_style]  (v5) edge (v6);
    \draw[edge_style]  (v6) edge (v7);
    \draw[edge_style]  (v7) edge (v8);
    \draw[edge_style]  (v5) edge (v9);

    \draw[edge_style]  (v6) edge (v9);
    \draw[edge_style]  (v7) edge (v10);
    \draw[edge_style]  (v8) edge (v10);
    \draw[edge_style]  (v9) edge (v10);
    \draw[edge_style]  (v5) to [loop left, in=120,out=240, looseness=10] (v5);
    \draw[edge_style]  (v8) to [loop right, in=-60,out=60,looseness=10 ] (v8);

    \draw[edge_styler] (v6) edge (v2);
    \draw[edge_styler] (v7) edge (v4);
    \draw[edge_styler] (v3) edge [bend left] (v1);
    \draw[edge_styler] (v9) edge [bend right] (v5);
    \draw[edge_styler] (v10) edge [bend left] (v8);

    \end{tikzpicture}
\end{subfigure}
\quad \quad 
\begin{subfigure}[b]{0.12\textwidth}
    \begin{tikzpicture}[scale=.3, shorten >=1pt, auto, node distance=1cm, ultra thick]
    \tikzstyle{node_style} = [circle,draw=black, inner sep=0pt, minimum size=4pt]
    \tikzstyle{edge_style} = [draw=black, line width=2, ultra thick]
    \tikzstyle{edge_styler} = [-,draw=red, line width=2, ultra thick, dashed]
    \tikzstyle{edge_styleg} = [-,draw=blue, line width=2, ultra thick, densely dotted]
    
    \node[node_style] (v1) at (0,0) {};
    \node[node_style] (v2) at (-2,2) {};
    \node[node_style] (v3) at (0,2) {};
    \node[node_style] (v4) at (2,2) {};
    \node[node_style] (v5) at (-3,4) {};
    \node[node_style] (v6) at (-1,4) {};
    \node[node_style] (v7) at (1,4) {};
    \node[node_style] (v8) at (3,4) {};
    \node[node_style] (v9) at (-1,6) {};
    \node[node_style] (v10) at (1,6) {};

    \draw[edge_styleg]  (v1) edge (v2);
    \draw[edge_style]  (v1) edge (v3);
    \draw[edge_style]  (v1) edge (v4);
    \draw[edge_style]  (v2) edge (v3);
    \draw[edge_styleg]  (v3) edge (v4);
    \draw[edge_style]  (v2) edge (v5);
    \draw[edge_style]  (v4) edge (v8);
    \draw[edge_style]  (v5) edge (v6);
    \draw[edge_styleg]  (v6) edge (v7);
    \draw[edge_style]  (v7) edge (v8);
    \draw[edge_style]  (v5) edge (v9);

    \draw[edge_style]  (v6) edge (v9);
    \draw[edge_style]  (v7) edge (v10);
    \draw[edge_style]  (v8) edge (v10);
    \draw[edge_styleg]  (v9) edge (v10);
    \draw[edge_styleg]  (v5) to [loop left, in=120,out=240, looseness=10] (v5);
    \draw[edge_styleg]  (v8) to [loop right, in=-60,out=60,looseness=10 ] (v8);

    \draw[edge_styler] (v6) edge (v2);
    \draw[edge_styler] (v7) edge (v4);
    \draw[edge_styler] (v3) edge [bend left] (v1);
    \draw[edge_styler] (v9) edge [bend right] (v5);
    \draw[edge_styler] (v10) edge [bend left] (v8);

    \end{tikzpicture}
    \end{subfigure} 
\\
\caption{\label{fig:graph}Left: The input graph $G$. Middle: The input graph and the edge set $A([n/2])$ marked dashed red. Note that any pairing of the vertices will do in the definition of the set. Right: A perfect matching $E'$ in the graph $G$ marked dotted blue. Note that the blue dotted and the red dashed edges together describe a $A([n/2])$-tangle that traverses all of $A([n/2])$.}
\end{figure}
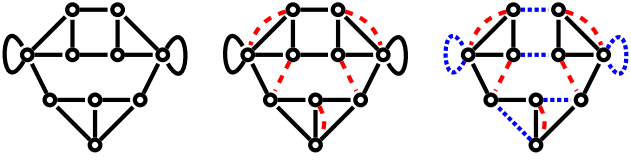
For a subset $Z\subseteq [n/2]$, define the edge set of the tangle 
\begin{align}
A(Z)=\{(w_{2i},w_{2i+1}): i\in Z\}. 
\end{align}
Let $G_Z$ be the input graph $G$ with the edges $A(Z)$ added to it. The algorithm uses the following inclusion--exclusion formula for the perfect matchings $M(G)$:
\begin{subequations}
\label{eq: main}
\begin{align}
M(G)&=\sum_{Z \in P([n/2])} (-1)^{n/2-|Z|} f_G(Z) \label{Pset},
\\
f_G(Z) &= \#\{a: a \mbox{ is an } A(Z)-\mbox{tangle in } G_Z\}.
\end{align}
\end{subequations}
In Eq. (\ref{Pset}) $P([n/2])$ denotes the set of all the subsets of $[n/2]=\{1,2,3,\ldots,n/2\}$ and $\#\{x: Q(x)\}$ indicates the number of $x$s that satisfy the clause $Q(x)$. Note that for a set of cardinality $n/2$ there are $2^{n/2}$ subsets. We will show in the next section that the function $f_G(Z)$ is polynomial time computable.

Let us argue now the correctness of Eq. (\ref{eq: main}). First consider a perfect matching $M$ in $G$. Note that $M\cup A([n/2])$ is exactly the edge set of an $A([n/2])$-tangle in $G_{[n/2]}$, cf. Fig. \ref{fig:graph}.
That is, the matchings together with the added alternating edges form even length cycles and paths with loops at both ends.
This $A([n/2])$-tangle will only be counted once in Eq. (\ref{eq: main}), namely for $Z=[n/2]$. It will not be counted for any other $Z$ as it traverses every edge in
$A([n/2])$ and at least one of them is missing in $G_Z$ for $Z\subset [n/2]$.

Second, any $A([n/2])$-tangle that traverses all of $A([n/2])$ in $G_{[n/2]}$ represents a unique perfect matching. This is because if one removes the edges in $A([n/2])$ one is again left with a perfect matching in $G$. So the equation at least counts perfect matchings, but we also must convince ourselves that it does not overcount.

To this end, consider a $A([n/2])$-tangle $\tau$ that does not traverse all edges in $A([n/2])$, say it only traverses $A(Y)$ for some $Y \subset [n/2]$. Then $\tau$ will be counted once in $G_{Z}$ for all $Z,Y\subseteq Z \subseteq [n/2]$, but with sign $(-1)^{n/2-|Z|}$. We have
\begin{align}
\sum_{Y\subseteq Z\subseteq [n/2]} (-1)^{n/2-|Z|}=\sum_{W\subseteq [n/2]\setminus Y}(-1)^{|W|}=0,
\end{align}
since there are as many odd as even sized subsets of a finite non-empty set. Hence, such a $\tau$ will not be counted in Eq. (\ref{eq: main}). \hfill\ensuremath{\square}

\subsection{Calculating $f_G(Z)$ in cubic time}\label{cubic-time}
In the last section we showed that the calculation of the hafnian boils down to the calculation of the function $f_G(Z)$ with $Z \in P([n/2])$.
Before showing how this is done let us introduce some additional notation.
For every adjacency matrix $\bm{A}$ we introduce its column-swapped version
\begin{align}\label{matA}
\bm{ \tilde{A}} &= \bm{X} \bm{A}, \text{ where}\\
\bm{X}&= \bigoplus_{i=1}^{n/2} \bm{\sigma}_X, \quad \bm{\sigma}_X = \left[ \begin{array}{cc}
0 & 1 \\
1 & 0
\end{array} \right].
\end{align}
{where we use the notation $\bigoplus$ to indicate the direct sum of matrices.} We label the submatrices of $\bm{\tilde{A}}$ by the set $Z=\{i_0,\ldots,i_{m-1}\} \subseteq [n/2]$ by using the notation $\bm{\tilde{A}}^{(Z)}$ to denote the $2m \times 2m$ square matrix obtained from $\bm{\tilde{A}}$ by keeping only rows and columns $\{ 2i_0,2i_0+1\ldots,2 i_{m-1}, 2 i_{m-1}+1\}$ from the original matrix $\bm{\tilde{A}}$.

To calculate $f_G(Z)$ we argue as follows. First count $A(Z)$-alternating walks by length, by looking at entries of matrix powers of a matrix obtained from the adjacency matrix $\bm{A}$ of $G$, keeping only the rows and columns representing the vertices spanned by $A(Z)$, and swapping every pair of columns that are connected in $A(Z)$; this matrix is precisely $\bm{\tilde{A}}^{(Z)}$. To show this note that the $(i,j)$ entry of $\bm{\tilde{A}}^{(Z)}$ carries the weight of walking from vertex $i$ to vertex $j$ using one original edge, and one ``red-dashed'' edge (cf. Fig. \ref{fig:graph}) i.e., an edge added by the edge set $A(Z)$. The trace of $(\bm{\tilde{A}}^{(Z)})^k$ counts closed alternating walks of length $2k$ ($k$ original edges and $k$ red ones), but it counts each walk $2k$ times.
The generating function
\begin{align}\label{poly}
p(\lambda , \bm{B}) = \sum_{j=1}^{n/2} \frac{1}{j!}\left(\sum_{k=1}^{n/2} \frac{\text{tr}(\bm{B}^k) \lambda^k}{(2k)} \right)^j.
\end{align}
counts in the monomial $\lambda^{n/2}$ the number of ways to combine several walks to a total of $n$ visited edges ($n/2$ original ones and $n/2$ red ones). Some edges may be counted multiple times here.
Finally using an inclusion-exclusion argument \cite{husfeldt2011invitation} it is seen that only combinations of walks that do not use the same edge twice survive in the summation (and there is precisely one such combination of closed alternating walks associated to each perfect matching in the original graph).

One can calculate all the traces appearing in Eq. (\ref{poly}) in cubic time by noting that if one uses the Schur decomposition 
\begin{align}\label{schur}
\bm{B} = \bm{Q} \bm{\Lambda} \bm{Q}^{-1},
\end{align} 
with $\bm{\Lambda}$ upper triangular, then 
\begin{align}\label{powertrace}
\text{tr}(\bm{B}^k) = \sum_{i} \Lambda_{i,i}^k.
\end{align}
{We use this decomposition because, in general, the matrix $\bm{B}$ is not diagonalizable (it is certainly not normal and thus the spectral theorem does not apply)\cite{horn1990matrix}, however the Schur decomposition is guaranteed to exist for any square matrix and is sufficient to calculate the power traces in Eq. (\ref{powertrace}) in cubic time in the size of the matrix using standard linear algebra routines \cite{anderson1999lapack}.} {Note that the Schur decomposition can be performed to very high accuracy but it is not rigorously exact since there is no analytical form for the roots of a polynomial of degree 5 or higher. In Appendix \ref{app:fastpol} we show that, in principle, the complexity of calculating the power traces can be reduced to $n^{\omega}$, without requiring any matrix diagonalization and requiring only additions and multiplications in the complex or real numbers. The quantity $2 \leq \omega \leq 3$ determines the number of operations required to do matrix multiplication of two square matrices of size $n$.}

All the quantities appearing in Eq. (\ref{poly}) are well defined not only for (0,1) matrices but for matrices over any field. This allows us to write the following formula for the hafnian of an arbitrary matrix $\bm{A}$ {by plugging Eq. (\ref{poly}) into Eq. (\ref{eq: main})}
\begin{align}\label{haffinal}
\haf(\bm{A}) =& \sum_{Z \in P([n/2])} (-1)^{n/2-|Z|} 
\times \\
& \quad  \left.\frac{1}{(n/2)!} \frac{d^{n/2}}{d\lambda^{n/2}} p\left(\lambda , \bm{\tilde{A}}^{(Z)}\right) \right|_{\lambda=0} \nonumber ,
\end{align}
and also for the loop hafnian
\begin{align}
\lhaf(\bm{A}) = &\sum_{Z \in P([n/2])} (-1)^{n/2-|Z|} \\
& \quad  \left.\frac{1}{(n/2)!} \frac{d^{n/2}}{d\lambda^{n/2}} q\left(\lambda, \bm{\tilde{A}}^{(Z)} , \text{diag}(\bm{A}^{(Z)}) \right) \right|_{\lambda=0} \nonumber ,
\end{align}
where now we have
\begin{align}
q(\lambda ,  \bm{B}, \bm{v} ) =& \sum_{j=1}^{n/2} \frac{1}{j!} \times \\
& \left(\sum_{k=1}^{n/2}  \left( \frac{\text{tr}(\bm{B}^k) }{(2k)} +\frac{\bm{v} (\bm{X} \bm{B})^{k-1} \bm{v}^T}{2} \right) \lambda^k \right)^j. \nonumber
\end{align}
The function $\text{diag}(\bm{B})$ returns the diagonal components of the matrix $\bm{B}$ as a row vector. If $\bm{v} = 0$ then $q(\lambda ,  \bm{B}, 0 ) =  p(\lambda ,\bm{B})$ and thus the lhaf reduces to the hafnian when the diagonal entries of the input matrix are zero.
Also note that 
\begin{align}
q(\lambda ,  \mu \bm{B}, \sqrt{\mu} \bm{v} ) = q(\mu \lambda, \bm{B}, \bm{v}),
\end{align} 
for any constant $\mu \geq 0$. This last equation shows explicitly that our loop hafnian formula conforms to the scaling relation in Eq. (\ref{scaling}). {Moreover, this formula shows interesting connections to generating functions for the permanent, determinant and the $\alpha-$permanents via the MacMahon Master Theorem \cite{quesada2018gaussian,lu2001macmahon,konvalinka2006non} 
}

\section{Numerical implementation and benchmarking}
We now discuss the results of our numerical implementation of the algorithms discussed in the previous section. 
For increased efficiency, we developed a C-programming-language--CPU-based version of the algorithm for benchmarking together with Python wrappers and also a sample implementation using Octave/Matlab\cite{hafnian}. This library will also be integrated in a future release of the Strawberry Fields platform \cite{killoran2018strawberry} for ease of use when studying Gaussian Boson Sampling.

We will consider three different types of graphs to benchmark the accuracy and the speed of the numerical computations: 
\begin{enumerate}
	\item Complete graphs with $n$ vertices: the hafnian of a complete graph, where all vertices are connected to one another with weight 1, without and with loops are known analytically and given by Eqs. (\ref{eq:haf1}) and (\ref{eq:haf2}) respectively. 
	
	\item Complete bipartite graphs with $n/2$ vertices: If we set the matrix $\bm{W}$ to have matrix elements $W_{i,j}=1$ in Eq. (\ref{eq:bipartite}), the hafnian of the matrix on the left side is simply ${\rm{per}}(\bm{W}) = (n/2)!$. 
	
	\item In order to test the speed of computations and go beyond the analytically known results, we will consider random symmetric matrices of size $n\times n$.
\end{enumerate}

For the first two sets of matrices the value of the hafnian is known, hence we used them to not only benchmark the speed of our implementation but also the numerical accuracy of the algorithm.

\subsection{Numerical implementation of the algorithm}

Numerical computations are performed using the Titan supercomputer which allows us to take advantage of hybrid CPU parallelism by combining distributed and shared memory methods using MPI and OpenMP protocols. The Titan supercomputer based at Oak Ridge National Laboratory has a theoretical peak performance of 27 petaFLOPS.\footnote{https://www.olcf.ornl.gov/olcf-resources/compute-systems/titan/} It has a Cray architecture and currently ranks among the top 5 supercomputers in the world. Further enhancements to our implementation can be made by using GPUs; this will require new implementations of batched, fast GPU based linear algebra routines for small matrices and will be the subject of a future study.

{As discussed in the previous section, computing the hafnian boils down to the following two steps: (i) evaluating the function $f_G(Z)$ for each $Z$ in the power set $P([n/2])$, and (ii) performing the outer summation in Eq. (\ref{Pset}).}

{The strength of the algorithm presented lies in the first step i.e. evaluating $f_G(Z)$ in cubic time. As described in the previous section, using Eq. (\ref{poly}) $f_G(Z)$ can be written in terms of the eigenvalues of the submatrices of the matrix $ \tilde{\bm{A}}$ in Eq. (\ref{matA}). This relation is exact (without any approximations involved) and is responsible for the $n^3$ contribution to the complexity of the algorithm as opposed to the $n^4$ factor in previous Hafnian algorithms. We employ LAPACK in order to compute the eigenvalues of the submatrices. Since these submatrices are often small in size, we evaluate $f_G(Z)$ for a given $Z$ in serial. While the algorithm itself is exact, numerical errors arise from the fact that LAPACK is available with at most complex double precision. This is one of the sources of numerical errors.}

{Note that in the second step i.e. evaluating the outer sum, for a matrix of size $n$ there are $2^{n/2}$ summands which is responsible for the exponential contribution to the complexity. However, since each of the summands can be evaluated independently of one another, we can utilize both distributed and shared memory CPU parallelism using MPI and OpenMP. The summation is distributed over multiple MPI nodes each having local OpenMP threads. First, partial sums are performed at each MPI node using multithreading and then at the end these partial sums are collected to the head node using \texttt{MPI sum reduction}. Thus, there is essentially no interprocess communication during the computation with reduction required only at the end. Due to this observation the Hafnian computation is embarrassingly parallel and scales very well with number of processors.}

\subsection{Results}
\begin{figure*}[tbh!]
	\includegraphics[width=0.47\textwidth]{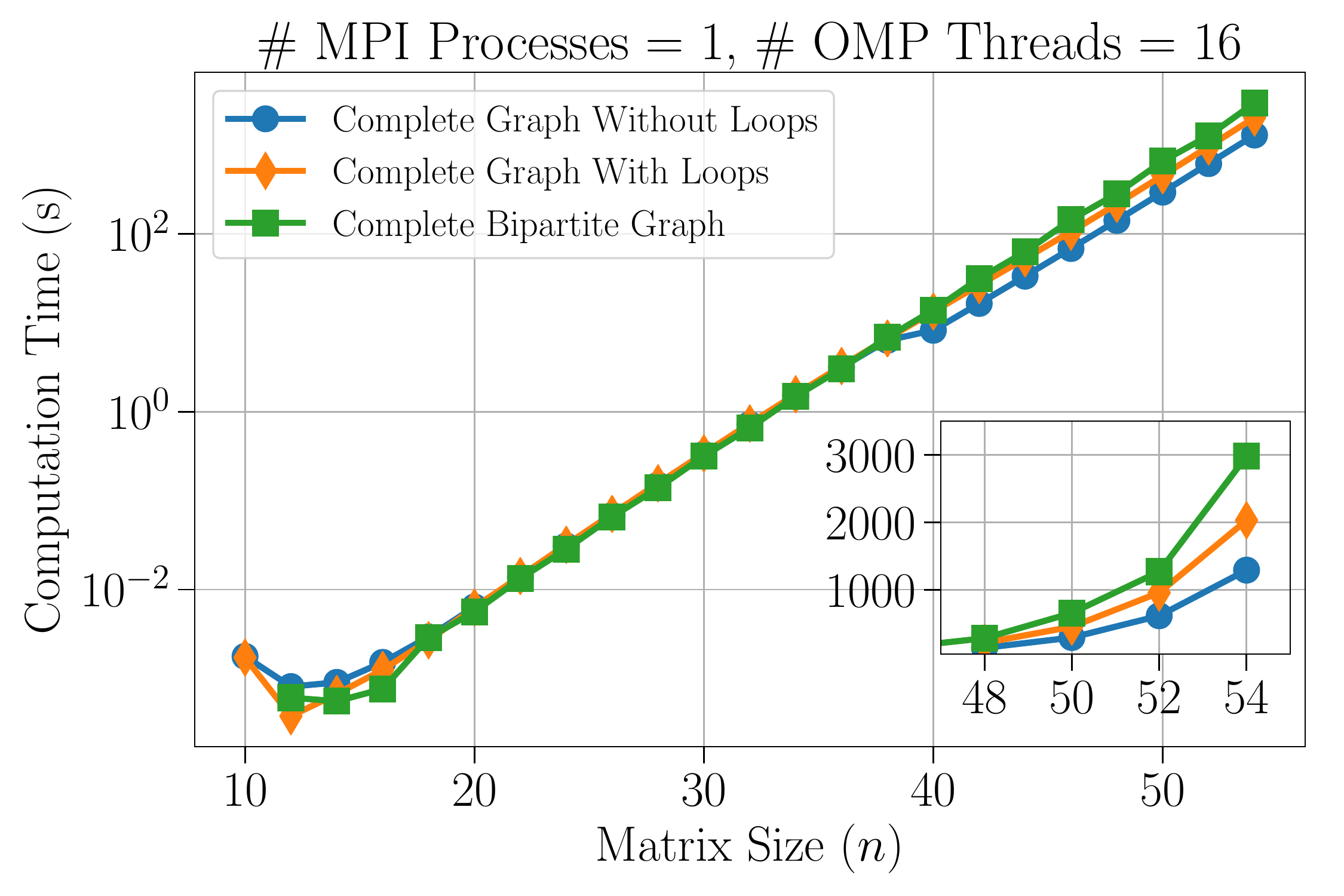}
	\
	\includegraphics[width=0.47\textwidth]{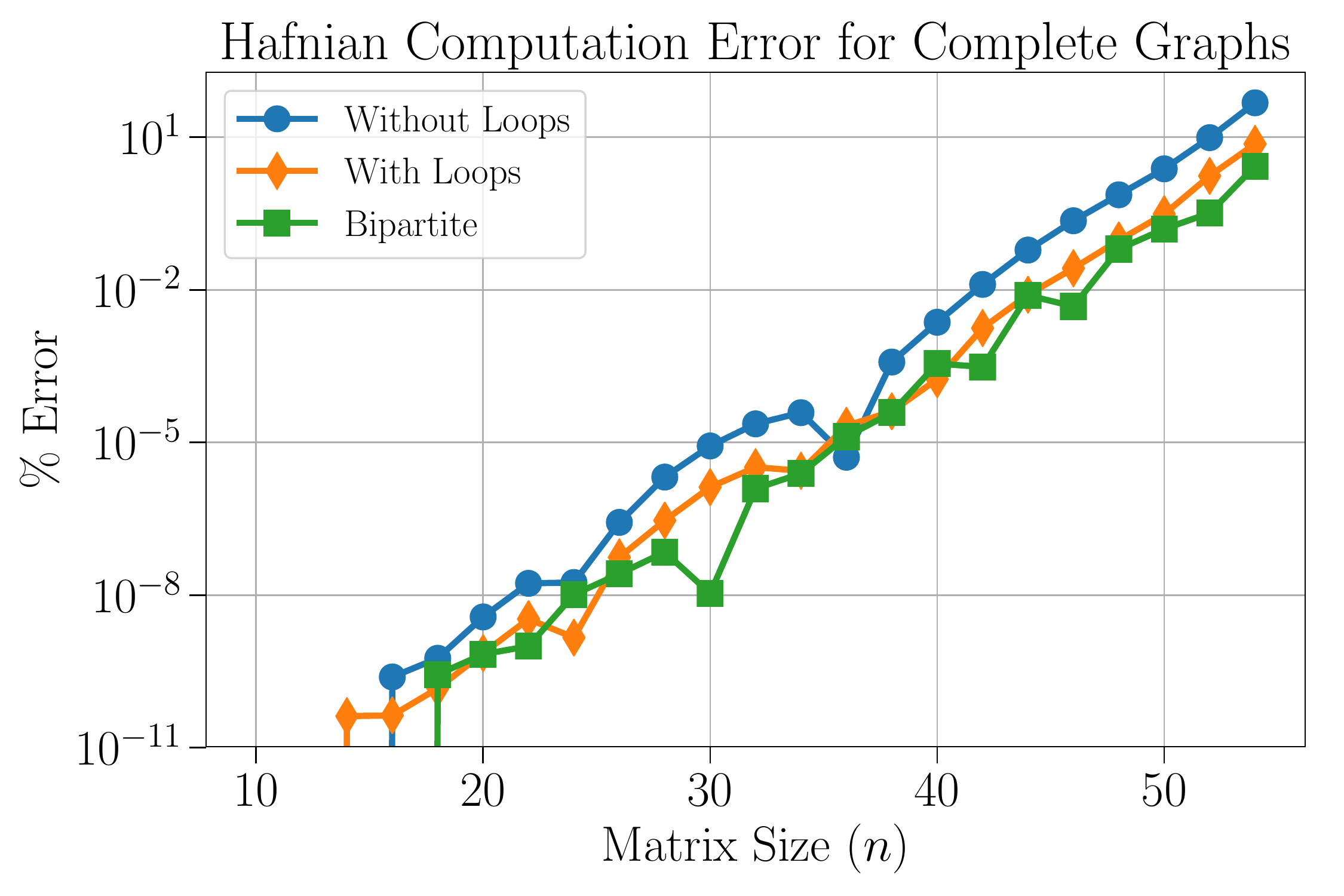}
	\caption{Hafnian computation time in seconds (left panel) and percentage error (right panel) plotted for various matrix sizes using single Message Passing Interface (MPI) process and 16 OpenMP threads for loop parallelism. It is evident that the computation time scales exponentially with $n$ and so does the fractional error defined $(\haf_{\text{numerical}}(\bm{M}) - \haf_{\text{exact}}(\bm{M}))/\haf_{\text{exact}}(\bm{M})$. For these benchmarks we used three different types of graphs for which the number of perfect matchings ($\haf$ or $\lhaf$) is known as a function of the matrix size $n$, complete graphs ($(n-1)!!$), complete graphs with loops ($T(n,-\tfrac{1}{2},\tfrac{1}{\sqrt{2}} $) and complete bipartite graphs ($n!$).}
	\label{fig:benchmark_n}
\end{figure*}
Let us now discuss the results of the numerical implementation of the algorithm. Fig. \ref{fig:benchmark_n} shows the performance benchmarks for complete and bipartite graphs by varying the size of the matrix (i.e., the number of vertices in the associated graph), using a single MPI process and 16 OpenMP threads for shared memory parallelism. The left panel shows the total computation time in seconds and the right panel shows the scaling of the percentage error defined as
\begin{align}\label{eq:error}
\% {\rm {Error}} = \frac{\haf_{\text{numerical}}(\bm{M}) - \haf_{\text{exact}}(\bm{M})}{\haf_{\text{exact}}(\bm{M})}\times 100,
\end{align}
where $\haf_{\text{numerical}}$ and $\haf_{\text{exact}}$ refer to numerical and analytical results respectively. 

The computation time scales exponentially with matrix size $n$ (the plots have logscale on the vertical axis) for all types of graphs including complete and bipartite graphs. As shown in the inset, for a complete graph without loops of size $n=54$ it takes approximately 1000 sec. For a graph with loops and a bipartite graph of the same size, the computation times are 2000 sec and 3000 sec respectively. Note that the exponential behaviour is only apparent for $n>20$. For small $n$ the program spends more time in preprocessing and setting up the computation which is responsible for a knee-like behaviour around $n=16$. By fitting the time scaling with the function $a~n^b 2^{cn}$ for $n>20$, we obtain $b\approx 3$ and $c\approx 1/2$ which is the expected scaling behaviour. The overall prefactor $a\approx 3.1\times 10^{-8}$ sec.

\begin{figure}[!ht]
	\includegraphics[width=0.45\textwidth]{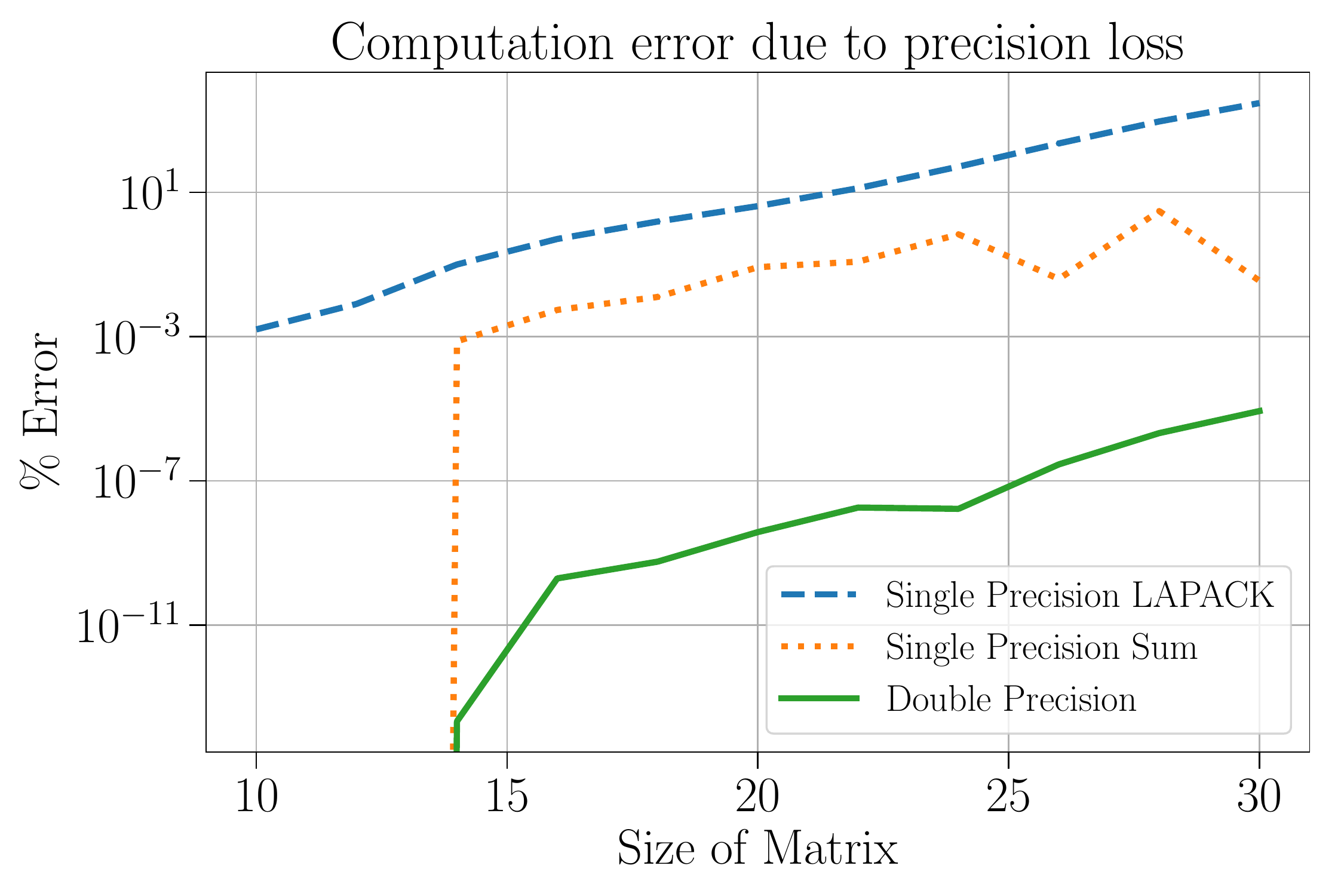}
	\caption{Comparison of Hafnian computation error due to using single precision for LAPACK and performing the summation. Clearly, the error due to low precision LAPACK dominates over the summation error. This indicates that the errors can further be controlled by employing quad precision LAPACK and summation.}
	\label{fig:singleprecision}
\end{figure}

Note that to compute the Hafnian function we need to perform $ \sim n^3 2^{n/2}$ floating point operations and each operation is associated with a small numerical error. We use the standard LAPACK linear algebra package for the computation of the eigenvalues of the submatrices appearing in Eq. (\ref{haffinal}), which is limited to {at most} double precision. As $n$ grows, the number of operations scales exponentially and so does the numerical error in the computation. The right panel of Fig. \ref{fig:benchmark_n} shows the scaling of the percentage error in the computation (defined as in Eq. \ref{eq:error}). For $n=54$, the number of operations is  $\sim 4.2\times10^{12}$, which comes with an error as large as $\sim 50\%$ for a complete graph without loops, $10\%$ for a graph with loops and $5\%$ for bipartite graphs. 

{There are two main sources of numerical errors: (i) limited precision of LAPACK, (ii) loss of precision in the numerical sum of an exponentially large number of elements. We distinguish these sources in Fig. \ref{fig:singleprecision}, by comparing the errors in the Hafnian calculation for a complete graph in the following three settings: (i) Using single precision LAPACK followed by casting the results to double precision and performing the sum in double precision (blue dashed curve), (ii) Using LAPACK in double precision followed by casting the results to single precision and performing the sum in single precision (orange dotted curve), and (iii) Using double precision for both LAPACK and the sum (green solid curve). Clearly, the error due to low precision in LAPACK dominates the low precision summation error. When both tasks are performed in double precision (in the third setting), errors are the lowest. This indicates that the computation errors can be controlled much better if quadruple precision is used for both tasks.}

\begin{figure}[!ht]
	\includegraphics[width=0.47\textwidth]{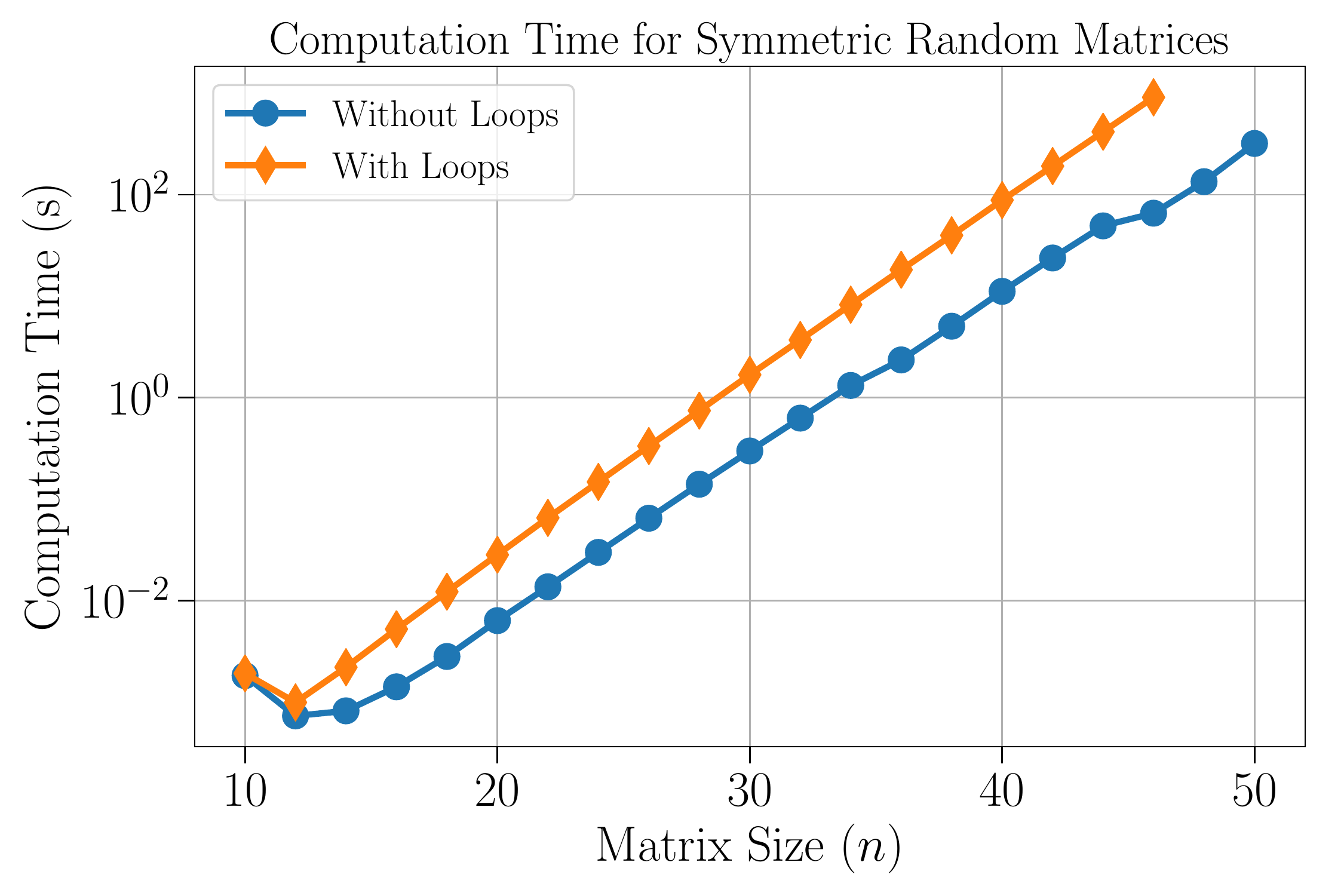}
	\caption{Hafnian computation time in seconds plotted for random symmetric matrices of various sizes using single MPI process and 16 OpenMP threads for loop parallelism. It is evident that similarly to Fig. \ref{fig:benchmark_n} computation time scales exponentially with $n$.}
	\label{fig:hafrandom}
\end{figure}

To test the scaling of the computation time, we also consider complex symmetric random matrices of various sizes in Fig. \ref{fig:hafrandom}. The real and imaginary parts of each element of the matrix are randomly chosen from a uniform distribution and then the resulting matrix is symmetrized. We find that similarly to the case of complete graphs (with matrix elements all being 1), the computation time scales exponentially with $n$. Moreover, since the loop hafnian function requires additional computations associated with the diagonal elements, the computation for those is larger as compared to the hafnian function. For a random matrix with $n=54$, the computation time is $\sim 4500$ sec and $1600$ sec for loop hafnian and hafnian respectively with 16 OpenMP threads. For a fixed size $n$ the computation times of random matrices are larger than those for matrices corresponding to complete graphs. This is because of the high symmetry of complete graphs which maps to a very simple structure of the eigenvalues of their adjacency matrices.

\begin{figure*}[!ht]
	\centering
	\includegraphics[width=0.47\textwidth]{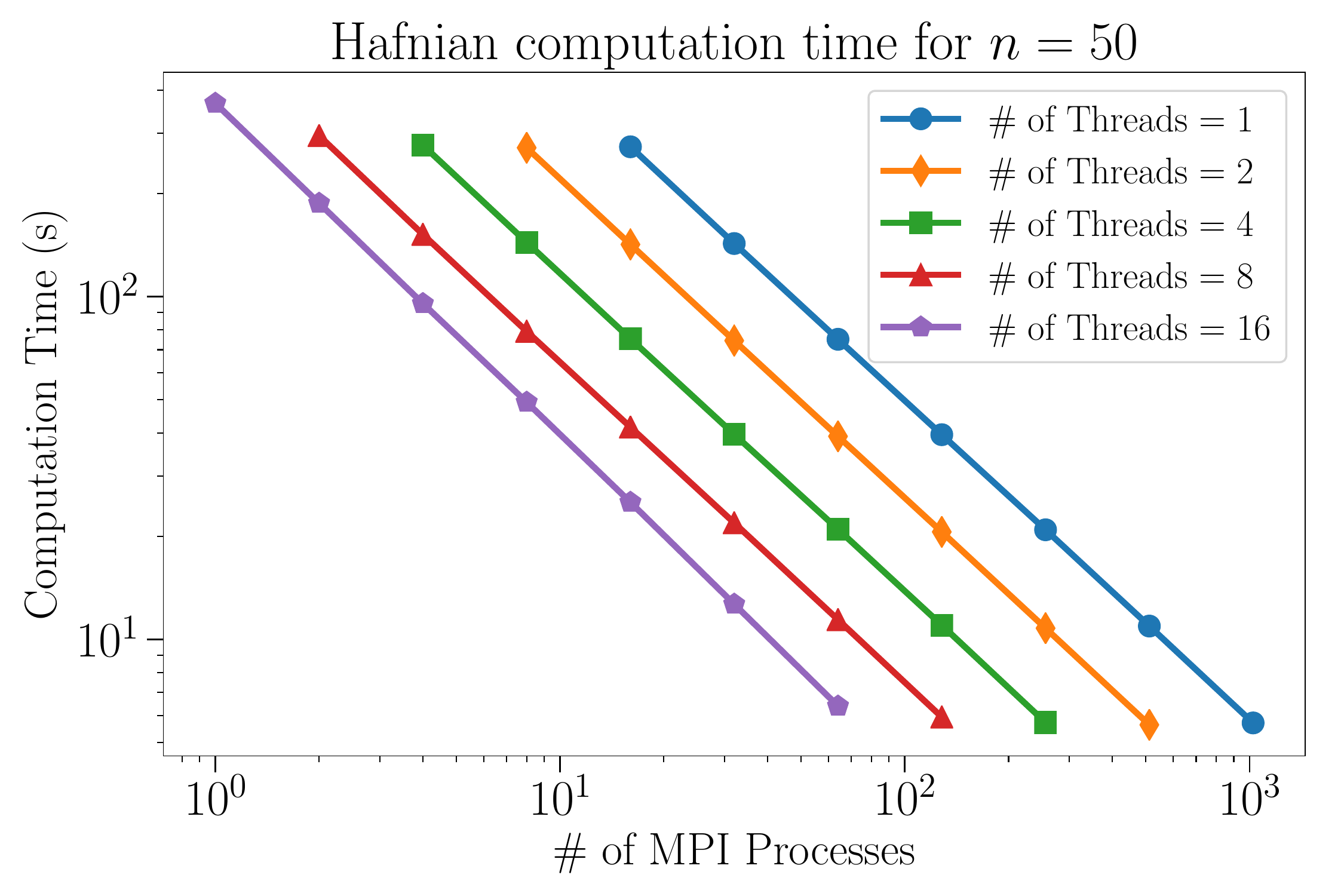}
	\
	\includegraphics[width=0.47\textwidth]{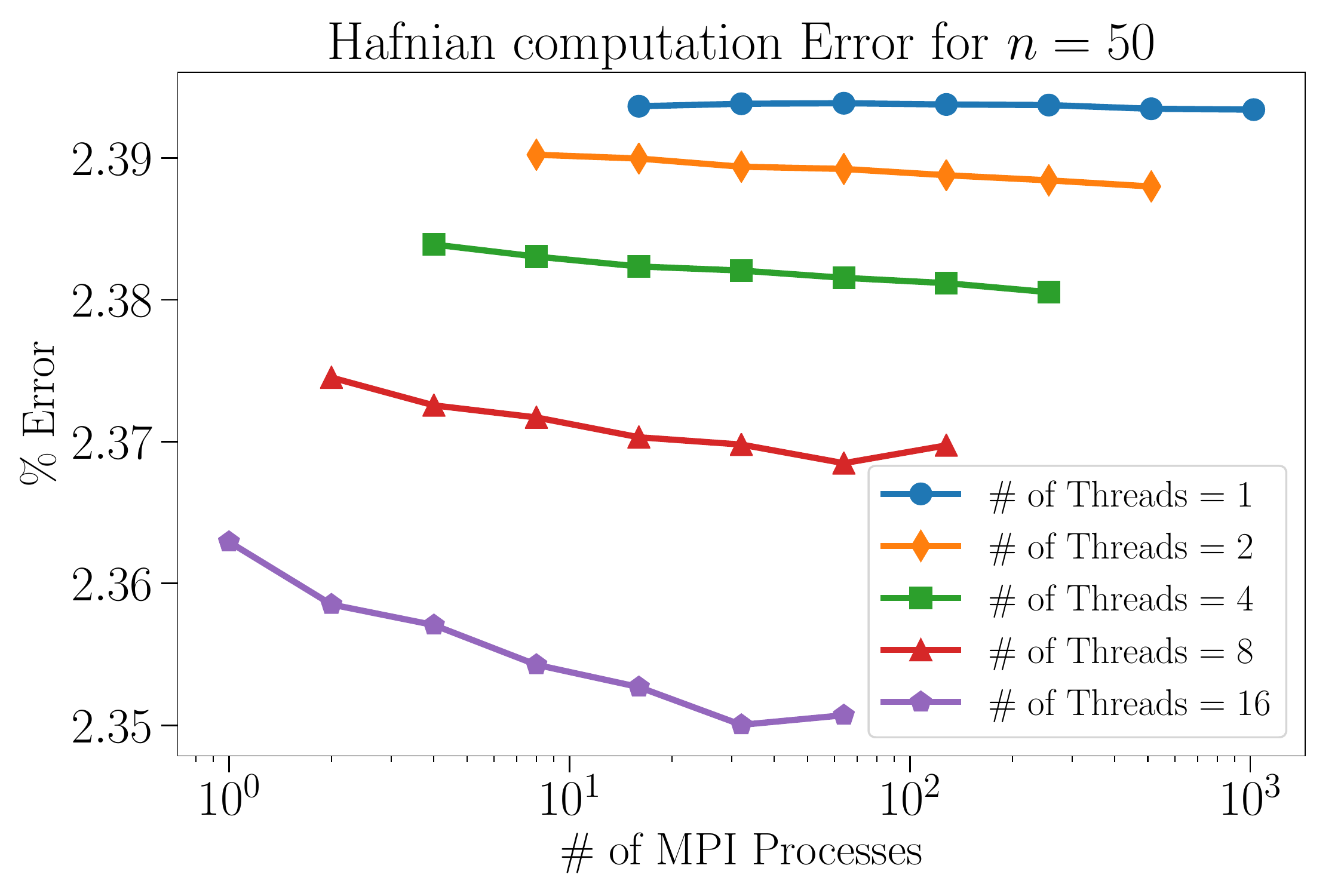}
	\caption{Scaling of the computation time (left panel) and percentage error (right panel) in hafnian computation for a complete graph with $n=50$ by considering various combinations of MPI and OpenMP processes. The left panel shows an almost perfect scaling with number of processors, while the percentage error (in the right panel) remains unchanged.}
	\label{fig:mpi}
\end{figure*}

\begin{figure}[!ht]
	\centering
	\includegraphics[width=0.45\textwidth]{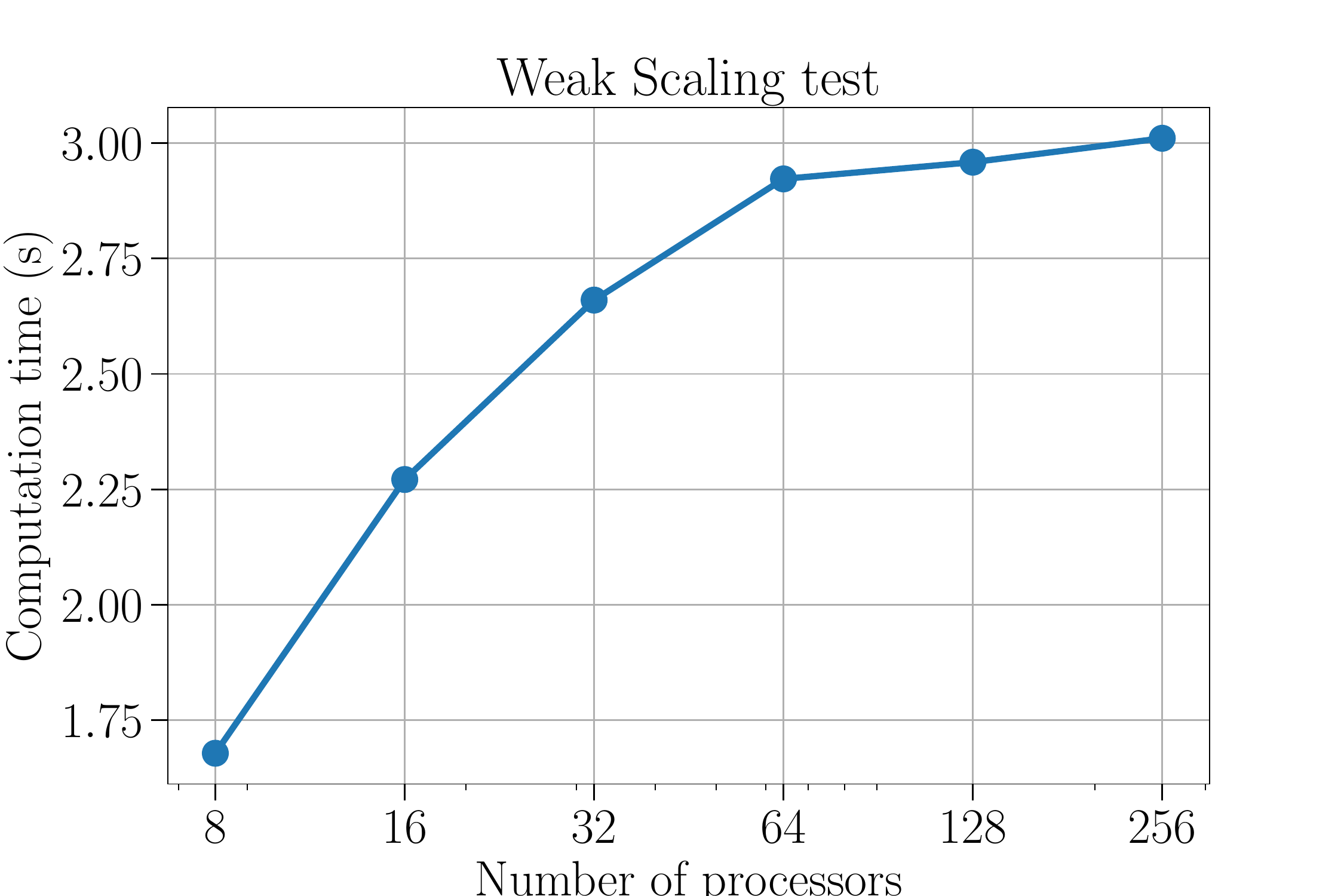}
	\caption{Weak scaling test performed by increasing the size of the matrix by 2 while doubling the number of processors each time. With increasing processor count the total time of computation approaches a constant value which indicates good weak scaling of the algorithm.}
	\label{fig:weakscaling}
\end{figure}

So far, we have only used shared memory parallelism on a single node computer using OpenMP multi-threading. We will now consider a hybrid CPU parallelism: distributing the computation over multiple nodes with OpenMP threads of their own. The time scaling using this hybrid approach for a matrix of size $n=50$ is shown in the left panel of Fig. \ref{fig:mpi}. We consider a range of MPI processes each with varying number of OpenMP threads. The computation turns out to perform extremely well with increasing number of MPI processes. The scaling is almost perfect, i.e. increasing MPI processes by a factor of 2 cuts down the computation time almost by a factor of half. Further enhancement in the computation time can be obtained by harnessing the power of GPU computing, which is beyond the scope of this implementation and is left for future work. {Note that numerical summation over multiple processors is a non-commutative process. Therefore, it is expected that when distributing the summation over multiple MPI and OpenMP processors can result in slightly different results. In order to test that, we compute  the associated percentage error in computation which is shown in the right panel of Fig. \ref{fig:mpi}.} It turns out that choosing different combinations of MPI and OpenMP processes have negligible effect on the accuracy of the results. Using GPUs or other vectorized computing techniques, one may be able to reduce the overall prefactor in the computation time scaling, but that leaves the exponential scaling shown in left panel of Fig. \ref{fig:benchmark_n} unaffected. {Fig. \ref{fig:weakscaling} shows the weak scaling test performed by doubling the number of MPI processes while increasing the size of the matrix by 2 each time. We consider matrix sizes $n = [32, 34, 36, 38, 40, 42]$ which corresponds to problem size $s = 2^{n/2}$ and the number of MPI processes considered $p =  [8,  16,  32,  64, 128, 256]$ each having 16 OpenMP threads. Hence, the ratio of the problem size to the number of processors is fixed to $s/p = 8192$. The total computation time plateaus as the processor count increases. This indicates a good weak scaling of the algorithm for larger processor count.}

\section{Summary}
From the discussion above, it is clear that evaluating hafnians of large matrices is limited by two factors: (i) speed and (ii) accuracy. For instance a hafnian computation would take $\sim 14000~{\rm sec} = 3.8~{\rm hours}$ for $n=60$ and $\sim 6.8\times10^{10}~{\rm sec} = 2155~{\rm years}$ for $n=100$ with 16 processors running in parallel. Utilizing all the 18000 CPU nodes (where each node has 16 CPUs, for a total of 288000 CPUs) of the Titan supercomputer and assuming perfect scaling over distributed nodes, computation of a single $n=100$ hafnian would take at least 1.5 months. This severely constrains the size of the problems where one needs to exactly compute many hafnians of large matrix sizes. For example, generating an (exact) sample for Gaussian Boson Sampling would likely require the evaluation of at least one hafnian of the size of the number of events that are sampled (i.e., the number of detectors that click). The above estimation shows that this problem becomes computationally intractable as the number of inputs on the linear interferometer is increased beyond a few tens. One can hope that for a problem of such extent, an ideal quantum device for Gaussian Boson Sampling may generate a sample in a much smaller time scale hence outperforming classical supercomputers. 

{As future work, it would be interesting to use GPUs as a way to speed up the calculation of hafnians. Progress in the bulk evaluation of the decompositions of real non-symmetric matrices has been presented by Tokura \emph{et al.} in Ref. \cite{tokura2017efficient}. However, for our purpose significant care is needed in sorting out the sizes of the different elements of the powerset sum required to evaluate the hafnian; moreover one would also need to extend the results of Tokura \emph{et al.} to complex non-hermitian matrices.}

\section{Acknowledgements}
B.G. and N.Q. thank J.M. Arrazola, J. Izaac, H. Jasim, N. Killoran, P. Rebentrost and C. Weedbrook for useful discussions and valuable feedback. This research used resources of the Oak Ridge Leadership Computing Facility at the Oak Ridge National Laboratory, which is supported by the Office of Science of the U.S. Department of Energy under Contract No. DE-AC05-00OR22725.

\bibliographystyle{unsrt}
\bibliography{hafnian}

\begin{thebibliography}{10}

\bibitem{barvinok2016combinatorics}
Alexander Barvinok.
\newblock {\em Combinatorics and complexity of partition functions}, volume
  274.
\newblock Springer, 2016.

\bibitem{hosoya1971topological}
Haruo Hosoya.
\newblock Topological index. a newly proposed quantity characterizing the
  topological nature of structural isomers of saturated hydrocarbons.
\newblock {\em Bulletin of the Chemical Society of Japan}, 44(9):2332--2339,
  1971.

\bibitem{quesada2018faster}
Nicol{\'a}s Quesada.
\newblock Franck-condon factors by counting perfect matchings of graphs with
  loops.
\newblock {\em J. Chem. Phys.}, 150(16):164113, 2019.

\bibitem{valiant1979complexity}
Leslie~G Valiant.
\newblock The complexity of computing the permanent.
\newblock {\em Theoretical computer science}, 8(2):189--201, 1979.

\bibitem{ryser1963combinatorial}
Herbert~John Ryser.
\newblock {\em Combinatorial mathematics}.
\newblock Number~14. Mathematical Association of America; distributed by Wiley
  [New York, 1963.

\bibitem{BaxFranklin}
E.~Bax and J.~Franklin.
\newblock A finite-difference sieve to compute the permanent.
\newblock {\em Caltech-CS-TR-96-04, California Institute of Technology}, 1996.

\bibitem{Balasubramanian}
K.~Balasubramanian.
\newblock Combinatorics and diagonals of matrices.
\newblock {\em Ph.D. Thesis, Department of Statistics, Loyola College, Madras,
  India, T073, Indian Statistical Institute, Calcutta}, 1980.

\bibitem{wu2016computing}
Junjie Wu, Yong Liu, Baida Zhang, Xianmin Jin, Yang Wang, Huiquan Wang, and
  Xuejun Yang.
\newblock A benchmark test of boson sampling on tianhe-2 supercomputer.
\newblock {\em National Science Review}, 5(5):715--720, 2018.

\bibitem{BarvinokIntro}
Alexander Barvinok.
\newblock Polynomial time algorithms to approximate permanents and mixed
  discriminants within a simply exponential factor.
\newblock {\em Random Structures \& Algorithms}, 14(1):29--61, 1999.

\bibitem{2016arXiv160107518B}
Alexander Barvinok.
\newblock Approximating permanents and hafnians.
\newblock {\em arXiv preprint arXiv:1601.07518}, 2016.

\bibitem{2014arXiv1409.3905R}
Mark Rudelson, Alex Samorodnitsky, Ofer Zeitouni, et~al.
\newblock Hafnians, perfect matchings and gaussian matrices.
\newblock {\em The Annals of Probability}, 44(4):2858--2888, 2016.

\bibitem{Chien:2004:DAC:982792.982903}
Steve Chien.
\newblock A determinant-based algorithm for counting perfect matchings in a
  general graph.
\newblock In {\em Proceedings of the Fifteenth Annual ACM-SIAM Symposium on
  Discrete Algorithms}, SODA '04, pages 728--735, Philadelphia, PA, USA, 2004.
  Society for Industrial and Applied Mathematics.

\bibitem{10.1007/3-540-36494-3_38}
Piotr Sankowski.
\newblock Alternative algorithms for counting all matchings in graphs.
\newblock In Helmut Alt and Michel Habib, editors, {\em STACS 2003}, pages
  427--438, Berlin, Heidelberg, 2003. Springer Berlin Heidelberg.

\bibitem{cygan2015faster}
Marek Cygan and Marcin Pilipczuk.
\newblock Faster exponential-time algorithms in graphs of bounded average
  degree.
\newblock {\em Information and Computation}, 243:75--85, 2015.

\bibitem{aaronson2011computational}
Scott Aaronson and Alex Arkhipov.
\newblock The computational complexity of linear optics.
\newblock In {\em Proceedings of the forty-third annual ACM symposium on Theory
  of computing}, pages 333--342. ACM, 2011.

\bibitem{neville2017classical}
Alex Neville, Chris Sparrow, Rapha{\"e}l Clifford, Eric Johnston, Patrick~M
  Birchall, Ashley Montanaro, and Anthony Laing.
\newblock Classical boson sampling algorithms with superior performance to
  near-term experiments.
\newblock {\em Nature Physics}, 13(12):1153, 2017.

\bibitem{clifford2018classical}
Peter Clifford and Rapha{\"e}l Clifford.
\newblock The classical complexity of boson sampling.
\newblock In {\em Proceedings of the Twenty-Ninth Annual ACM-SIAM Symposium on
  Discrete Algorithms}, pages 146--155. SIAM, 2018.

\bibitem{hamilton2017gaussian}
Craig~S Hamilton, Regina Kruse, Linda Sansoni, Sonja Barkhofen, Christine
  Silberhorn, and Igor Jex.
\newblock Gaussian boson sampling.
\newblock {\em Physical review letters}, 119(17):170501, 2017.

\bibitem{kruse2018detailed}
Regina Kruse, Craig~S Hamilton, Linda Sansoni, Sonja Barkhofen, Christine
  Silberhorn, and Igor Jex.
\newblock A detailed study of gaussian boson sampling.
\newblock {\em arXiv preprint arXiv:1801.07488}, 2018.

\bibitem{lvovsky2014squeezed}
AI~Lvovsky.
\newblock Squeezed light.
\newblock {\em Photonics Volume 1: Fundamentals of Photonics and Physics},
  pages 121--164, 2015.

\bibitem{B96}
Alexander~I Barvinok.
\newblock Two algorithmic results for the traveling salesman problem.
\newblock {\em Mathematics of Operations Research}, 21(1):65--84, 1996.

\bibitem{harrow2017quantum}
Aram~W Harrow and Ashley Montanaro.
\newblock Quantum computational supremacy.
\newblock {\em Nature}, 549(7671):203, 2017.

\bibitem{quesada2018gaussian}
Nicol\'as Quesada, Juan~Miguel Arrazola, and Nathan Killoran.
\newblock Gaussian boson sampling using threshold detectors.
\newblock {\em Phys. Rev. A}, 98:062322, 2018.

\bibitem{gupt2018classical}
Brajesh Gupt, Juan~Miguel Arrazola, Nicol{\'a}s Quesada, and Thomas~R Bromley.
\newblock Classical benchmarking of gaussian boson sampling on the titan
  supercomputer.
\newblock {\em arXiv preprint arXiv:1810.00900}, 2018.

\bibitem{bjorklund2008exact}
Andreas Bj{\"o}rklund and Thore Husfeldt.
\newblock Exact algorithms for exact satisfiability and number of perfect
  matchings.
\newblock {\em Algorithmica}, 52(2):226--249, 2008.

\bibitem{kan2008moments}
Raymond Kan.
\newblock From moments of sum to moments of product.
\newblock {\em Journal of Multivariate Analysis}, 99(3):542--554, 2008.

\bibitem{koivisto2009partitioning}
Mikko Koivisto.
\newblock Partitioning into sets of bounded cardinality.
\newblock In {\em International Workshop on Parameterized and Exact
  Computation}, pages 258--263. Springer, 2009.

\bibitem{nederlof2009fast}
Jesper Nederlof.
\newblock Fast polynomial-space algorithms using m{\"o}bius inversion:
  Improving on steiner tree and related problems.
\newblock In {\em International Colloquium on Automata, Languages, and
  Programming}, pages 713--725. Springer, 2009.

\bibitem{bjorklund2012counting}
Andreas Bj{\"o}rklund.
\newblock Counting perfect matchings as fast as ryser.
\newblock In {\em Proceedings of the twenty-third annual ACM-SIAM symposium on
  Discrete Algorithms}, pages 914--921. SIAM, 2012.

\bibitem{rempala2007symmetric}
Grzegorz Rempala and Jacek Wesolowski.
\newblock {\em Symmetric functionals on random matrices and random matchings
  problems}, volume 147.
\newblock Springer Science \& Business Media, 2007.

\bibitem{TAOCP}
Donald~E. Knuth.
\newblock {\em The Art of Computer Programming, Vol. 2: Seminumerical
  Algorithms}.
\newblock Addison-Wesley, 3 edition, 1998.

\bibitem{abramowitz1964handbook}
Milton Abramowitz and Irene~A Stegun.
\newblock {\em Handbook of mathematical functions: with formulas, graphs, and
  mathematical tables}, volume~55.
\newblock Dover, 1972.

\bibitem{husfeldt2011invitation}
Thore Husfeldt.
\newblock Invitation to algorithmic uses of inclusion--exclusion.
\newblock In {\em International Colloquium on Automata, Languages, and
  Programming}, pages 42--59. Springer, 2011.

\bibitem{horn1990matrix}
Roger~A Horn, Roger~A Horn, and Charles~R Johnson.
\newblock {\em Matrix analysis}.
\newblock Cambridge university press, 1990.

\bibitem{anderson1999lapack}
Edward Anderson, Zhaojun Bai, Christian Bischof, L~Susan Blackford, James
  Demmel, Jack Dongarra, Jeremy Du~Croz, Anne Greenbaum, Sven Hammarling, Alan
  McKenney, et~al.
\newblock {\em LAPACK Users' guide}.
\newblock SIAM, 1999.

\bibitem{lu2001macmahon}
I-Li Lu and Donald St~P Richards.
\newblock Macmahon's master theorem, representation theory, and moments of
  wishart distributions.
\newblock {\em Advances in Applied Mathematics}, 27(2-3):531--547, 2001.

\bibitem{konvalinka2006non}
Matjaz Konvalinka and Igor Pak.
\newblock Non-commutative extensions of the macmahon master theorem.
\newblock {\em arXiv preprint math/0607737}, 2006.

\bibitem{hafnian}
hafnian.
\newblock \url{https://github.com/XanaduAI/hafnian}, 2018.

\bibitem{killoran2018strawberry}
Nathan Killoran, Josh Izaac, Nicol{\'{a}}s Quesada, Ville Bergholm, Matthew
  Amy, and Christian Weedbrook.
\newblock Strawberry {F}ields: {A} {S}oftware {P}latform for {P}hotonic
  {Q}uantum {C}omputing.
\newblock {\em {Quantum}}, 3:129, 2019.

\bibitem{tokura2017efficient}
Hiroki Tokura, Takumi Honda, Yasuaki Ito, Koji Nakano, Mitsuya Nishino, Yushiro
  Hirota, and Masami Saeki.
\newblock An efficient gpu implementation of bulk computation of the eigenvalue
  problem for many small real non-symmetric matrices.
\newblock {\em International Journal of Networking and Computing},
  7(2):227--247, 2017.

\bibitem{storjohann1998n}
Arne Storjohann.
\newblock An o (n 3) algorithm for the frobenius normal form.
\newblock In {\em ISSAC}, volume~98, pages 101--104. Citeseer, 1998.

\bibitem{giesbrecht1995nearly}
Mark Giesbrecht.
\newblock Nearly optimal algorithms for canonical matrix forms.
\newblock {\em SIAM Journal on Computing}, 24(5):948--969, 1995.

\bibitem{keller1985fast}
Walter Keller-Gehrig.
\newblock Fast algorithms for the characteristics polynomial.
\newblock {\em Theoretical computer science}, 36:309--317, 1985.

\bibitem{neunhoffer2008computing}
Max Neunh{\"o}ffer and Cheryl~E Praeger.
\newblock Computing minimal polynomials of matrices.
\newblock {\em LMS Journal of Computation and Mathematics}, 11:252--279, 2008.

\end{thebibliography}
\appendix

\section{Number of elements in SPM($n$)}\label{countingSPM}
For $n$ even one can obtain a closed form expression for the number of elements in SPM($n$). 
Consider first PMP$(n)$  the set of perfect matchings in a graph with $n$ vertices. One can start breaking pairs, i.e., taking a matching such as $(0,1)$ and turn into two loops $(0,0)(1,1)$. If one ``breaks'' $j$ pairs there are $n/2 \choose j$ ways of doing this for each perfect matching. However this will overcount the number of partitions. For example if one breaks every matching of the following three perfect matchings $(0,1),(2,3)$ , $(0,2),(1,3)$ and $(0,3),(1,2)$ one will always get the same set of loops $(0,0),(1,1),(2,2),(3,3)$ thus one needs to account for multiple counting by dividing by the factor $(j-1)!!$. We conclude that if one breaks $j$ pairs from the set of perfect matching of $n$ objects one gets $\frac{(n-1)!!}{(j-1)!!} { n/2 \choose j},$ partitions. We now need to sum over all possible $j$ to obtain the total number of partitions
\begin{align}
|\text{SPM}(n)|&=\sum_{j=0}^{n/2} \frac{(n-1)!!}{(j-1)!!} { n/2 \choose j} = T\left(n,-\tfrac{1}{2},\tfrac{1}{\sqrt{2}}\right) \nonumber\\
&= (n-1)!! \ M\left(-\tfrac{n}{2},\tfrac{1}{2},-\tfrac{1}{2}\right) \nonumber \\
&= (n-1)!!  \ e^{-1/2} \  M\left(\tfrac{n+1}{2},\tfrac{1}{2},\tfrac{1}{2}\right),
\end{align}
where $T(a,b,r)$ is the Toronto function and $M\left(a,b,z\right)$ is the Kummer confluent hypergeometric function \cite{abramowitz1964handbook}.
We also have the following asymptotic limit and bound:
\begin{align}
&\lim_{n \to \infty} \frac{M\left(-\tfrac{n}{2},\tfrac{1}{2},-\tfrac{1}{2}\right)}{ \exp(\sqrt{n}-\tfrac{1}{4})/2} = 1,\\
& M\left(-\tfrac{n}{2},\tfrac{1}{2},-\tfrac{1}{2}\right) > \exp(\sqrt{n}-\tfrac{1}{4})/2.
\end{align}

\section{Fast calculation of power traces}\label{app:fastpol}
For the calculation of the hafnian of an $n \times n$ matrix it is required to know the quantities
\begin{align}
\text{tr}\left(\bm{B}^k\right), \quad 1 \leq k \leq n.
\end{align}
This can be done by first obtaining the minimal polynomial of the matrix $\bm{B} \in \mathbb{C}^{m \times m}$, which is the monic polynomial $p(x)$ of least degree such that $p(\bm{B})$ = 0. Note that for an $m \times m$ matrix the minimal polynomial is at most of degree $m$ by the Cayley-Hamilton theorem. Assuming that $k \leq m$ is the degree of the minimal polynomial we write it as
\begin{align}
p(x) = x^k+\sum_{j=0}^{k-1} c_j x^j,
\end{align}
and using the fact that $p$ is the minimal polynomial of $\bm{B}$ we write
\begin{align}
\bm{B}^{k} &= -\sum_{j=0}^{k-1} c_j \bm{B}^{j} \longrightarrow \\
\bm{B}^{k+l} &= -\sum_{j=0}^{k-1} c_j \bm{B}^{j+l} \longrightarrow \\
\text{tr}\left( \bm{B}^{k+l}\right) &= -\sum_{j=0}^{k-1} c_j \text{tr}\left( \bm{B}^{j+l}\right)
\end{align}
Thus, once the minimal polynomial is known, one can easily calculate any power trace of degree $n=k+l$ of a matrix. 
The minimal polynomial calculation can be done in time $O(k^\omega)$ where $2 \leq \omega \leq 3$ quantifies the complexity of matrix-matrix multiplication \cite{storjohann1998n,giesbrecht1995nearly,keller1985fast,neunhoffer2008computing}. Note that these algorithms only require operations within the field in which the matrix is expressed and are exact in infinite precision arithmetic. Yet, when used in finite precision arithmetic (i.e. in real hardware with finite RAM) these algorithms do not perform better in terms accuracy than a Schur factorization of the matrix followed by exponentiation and summation of the eigenvalues for calculating power traces. The same conclusion holds for the calculation of $f_G(Z)$ in Sec. \ref{cubic-time} and thus we prefer to use the Schur decomposition for the calculation of the power traces.

\section{Hafnians of low rank matrices}\label{app:lowrank}
{We argue how the hafnian of a matrix $\bm{A}\in \mathbb{K}^{2n\times 2n}$ that has rank $r$ over the field $\mathbb{K}$, can be computed in time $\binom{2n+r-1}{r-1}\operatorname{poly}(n)$. This generalizes a result of Barvinok for the permanent \cite{B96}.

Let $\mbox{\small PMP}(2n)$ be the set of perfect matching permutations on $2n$ elements, i.e., permutations $\sigma:[2n]\rightarrow [2n]$ such that
$\forall i:\sigma(2i-1)<\sigma(2i)$ and $\forall i:\sigma(2i-1)<\sigma(2i+1)$.

Given a matrix $\bm{A}\in \mathbb{K}^{2n\times 2n}$, its hafnian is
\begin{equation}
\operatorname{haf}(\bm{A})=\sum_{\sigma\in \mbox{\small PMP}(2n)} \prod_{i=1}^n A_{\sigma(2i-1),\sigma(2i)}.
\end{equation}
Note that for any lower triangular matrix $\bm{L}\in \bm{K}^{2n\times 2n}$,
\begin{equation}
\operatorname{haf}(\bm{A})=\operatorname{haf}(\bm{A}+\bm{L}),
\end{equation}
since the hafnian, as defined above, does only depend on elements above the diagonal. 

If we can find a $2n\times r$ matrix $\bm{G}$ for any lower triangular $\bm{L}$ such that
\begin{equation}
\bm{G}\bm{G}^{\mbox{\tiny T}}=\bm{A}+\bm{L},
\end{equation}
for a small enough $r$, there is a relatively efficient algorithm to compute the hafnian following the permanent algorithm by Barvinok~\cite{B96}, as follows:
Introduce $r$ indeterminates $x_1,\ldots,x_r$, and consider the multivariate polynomial
\begin{equation}
\label{eq: q}
q(x_1,\ldots,x_r)=\prod_{i=1}^{2n} \sum_{j=1}^r g_{i,j}x_j.
\end{equation}

Let $\mathcal{P}_{2n,r}$ be the set of integer $r$-partitions of $2n$, i.e., tuples $(p_1,p_2,\ldots,p_{r})$ such that
\begin{enumerate}
\item $\forall i:p_i\geq 0$,
\item $\sum_i p_i=2n$.
\end{enumerate}
Let $\mathcal{E}_{2n,r}$ be the subset of $\mathcal{P}_{2n,r}$ of \emph{even} partitions, i.e., tuples $(p_1,p_2,\ldots,p_{r})\in \mathcal{P}_{2n,r}$ such that also $p_i$ is even for all $i$.
Note that 
\begin{equation}
|\mathcal{P}_{2n,r}|=\binom{2n+r-1}{r-1} \qquad \mbox{and} \qquad |\mathcal{E}_{2n,r}|=\binom{n+r-1}{r-1}.
\end{equation}

Expand~(\ref{eq: q}) to identify the non-zero coefficients $\lambda_p$ in
\begin{equation}
\label{eq: qexp}
q(x_1,\ldots,x_r)=\sum_{p=(p_1,\ldots,p_{r})\in \mathcal{P}_{2n,r}} \lambda_p\prod_{i=1}^{r} x_i^{p_i}.
\end{equation}
This can be done in $|\mathcal{P}_{2n,r}|\operatorname{poly}(n)$ time.

Let $a!!$ denote the \emph{double factorial} of $a$, i.e., $\prod_{k=0}^{\lceil a/2 \rceil-1} (a-2k)$. We use the convention that $a!!$ for $a\leq 0$ is $1$. The hafnian can now be expressed in the coefficients of $q(x_1,\ldots,x_{r})$ as
\begin{equation}
\operatorname{haf}(\bm{A})=\sum_{e\in \mathcal{E}_{2n,r}} \lambda_e \prod_{i=1}^{r} (e_i-1)!!.
\end{equation}
Given the coefficients $\lambda_e$, this can be evaluated in $|\mathcal{E}_{2n,r}|\operatorname{poly}(n)$ time.
Note that since we only need the coefficients $\lambda_e$ for $e\in \mathcal{E}_{2n,r}$, it is interesting to investigate if there are faster ways to obtain these than computing all of~(\ref{eq: qexp}) explicitly.
}
\end{document}